# The arrow of time (second law) as a randomness-driven emergent property of large systems

**D. Wright, R. Klein-Seetharaman, and S. K. Sarkar**[*]

*Department of Physics, Colorado School of Mines, Golden, CO 80401, U.S.A.*

*Corresponding author:*
[*]ssarkar@mines.edu

**Abstract**

The arrow of time is an irreversible phenomenon for a system of particles undergoing reversible dynamics. Since the time of Boltzmann to this day, the arrow of time has led to debate and research. However, the enormous growth of nanotechnology and associated experimental techniques has brought the arrow of time at the forefront because of its practical implications. Using simulations of one-dimensional diffusion of a system of particles, we show that the arrow of time is an emergent property of a large system. We show that the recurrence time for a system of particles to return to its original configuration grows rapidly as the number of particles grows. Based on the simulations, we have provided the expressions for recurrence times for classical particles, Fermions, and Bosons. A system of Bosons has the shortest recurrence time, whereas a system of classical particles has the longest recurrence time. The underlying distribution around the mean recurrence time is Poisson-distributed for Bosons and Gaussian-distributed for Fermions and classical particles. The probabilistic approach to encode dynamics enables testing processes other than diffusion and quantify their effects on the recurrence time.

**Introduction**

The equation of motion describing the dynamics of a particle has time-reversal symmetry in both classical and quantum mechanics (1). In simple terms, an observer cannot distinguish the forward and reverse directions of a movie showing the dynamics of the particle, such as swinging of a pendulum. Despite the underlying time-reversal symmetry, a system of a large number of particles reveals an irreversible direction of time, coined as the arrow of time by Eddington (2). Boltzmann formulated this irreversibility as the second law of thermodynamics. The second law dictates that the entropy always increases for spontaneous processes (3). The arrow of time has led to confusion, debate, and research since the time of Boltzmann (4).

With the advent of single molecule techniques, research has shown that entropy can spontaneously decrease for small systems consisting of a single particle or a few particles (5). These results suggest that the arrow of time or the second law of thermodynamics may not be something fundamental. Instead, the arrow of time is an emergent property of large systems. In principle, a group of particles can go back to its initial configuration. However, the probability of going back to the initial state becomes exceedingly low as the system becomes large. Fluctuation theorems quantify the probabilities in both the forward and backward directions (6-9). Experiments have confirmed these fluctuation theorems (10-13).

The possibility of violations of the second law of thermodynamics in smaller spatial and temporal timescales have significant implications at the nanoscale (14), including electronics and biochemistry. As a result, there has been a substantial body of research on the impact of thermodynamics for smaller systems. To this end, we note that the defining feature for small dynamical systems is fluctuations or randomness (15). Entropy quantifies the randomness, and researchers have formulated different definitions of entropy suitable for different situations, including equilibrium and nonequilibrium conditions. Once the entropy is defined, experimentally measurable thermodynamic quantities follow. Recent work has even enabled calculation of equilibrium free energies from nonequilibrium experiments (16-18). Therefore, the breaking of the arrow of time or violations of the second law is an important contemporary topic of statistical mechanics and thermodynamics. However, most papers in this area are difficult for students' understanding of the matter. There is a lack of simple examples where one can understand the concept of the arrow of time.

In this paper, we simulate a system of random walkers in one dimension (1D) with a varying number of particles. Initially, we distributed the particles at random locations within the 1D grid of boxes. Particle statistics determined the rule of occupancy in each box. Consequently, each particle moved to the left or the right with equal probabilities. We analyzed the recurrence times and found that Bosons return to the initial position with the shortest average recurrence time, followed by Fermions and classical particles. Based on the simulations, we defined the analytical expressions of recurrence times for particles obeying classical, Fermi-Dirac, and Bose-Einstein statistics. The distribution of mean values obtained from repeats of simulations follows a gamma distribution for Bosons and a Gaussian distribution for Fermions and classical particles. The probabilistic approach enables the encoding of any type of motion for the particles. Also, one can encode both spatial and temporal heterogeneities to test the effects on recurrence. For example, we simulated a single Poisson process for the entire duration of simulations, but more than one Poisson process can describe different time intervals to simulate dynamic heterogeneities. The simple approach to study recurrence times will facilitate research and teaching under different simulation conditions.

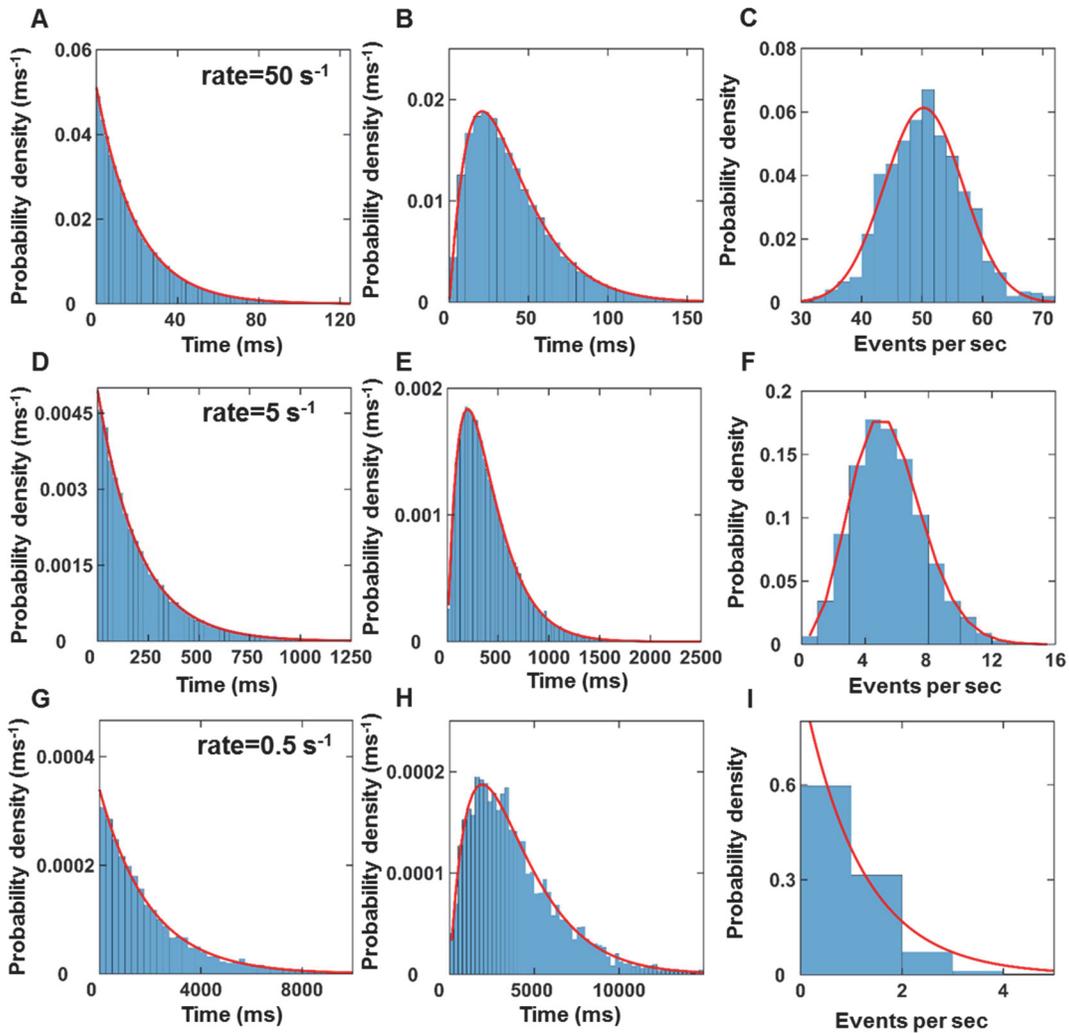

**Figure 1. Poisson process.** (**A**), (**D**), and (**G**) Area-normalized histograms of times between consecutive events. (**B**), (**E**), and (**H**) Area-normalized histograms of times between every second event for 50 s$^{-1}$, 5 s$^{-1}$, and 0.5 s$^{-1}$, respectively. (**C**), (**F**), and (**I**) Area-normalized histograms of events per second for a 50 s$^{-1}$, 5 s$^{-1}$, and 0.5 s$^{-1}$, respectively.

**Results and discussion**
**A Poisson process has well-defined consequences.** A Poisson process or a chain of Poisson processes describes many natural and random processes such as radioactive decay (19), cellular locations in tissue models (20), inter-domain dynamics of proteins (21), chemical reactions (22), and so on. A Poisson process has a constant probability of occurring at each temporal or spatial step. There are three consequences of a Poisson process: 1) the distribution of time steps between consecutive events is an exponential distribution, 2) the distribution of time steps between every second or more event is a gamma distribution, 3) the distribution of the number of events in an interval of time or space has a wide range of possible shapes including Gaussian distribution and exponential distribution. First, we simulated a Poisson process with a rate of $k = 50$ s$^{-1}$. We used a time step $dt = 1$ ms for simulations. The probability of such a Poisson process occurring at a time step $dt = 1\,\text{ms}$ is given by $P = P(t)\,dt = k \times \exp(-k \times dt) \times dt$, where $P(t)$ is the probability density function. For $k = 50\,\text{s}^{-1}$, $P = 0.0476$ for $dt = 1\,\text{ms}$. At each time step, we generated a random number between 0 and 1. If the random number produced satisfied the condition $N \leq 0.0476$, we assigned 1 at that time point; otherwise, we assigned 0. As a result, we generated a time series $\{0,0,1,0,0,1,1,0,0,...\}$ with $10^7$ time points. We noted the time between consecutive events. For example, the time series $\{0,0,1,0,0,1,1,0,0,...\}$ has three positive events. We can calculate two instances of the times between consecutive 1s: $\Delta t_1 = 3$ and $\Delta t_1 = 1$. We can also calculate one example of the time between every second event: $\Delta t_2 = 4$.

We made a list of all $\Delta t$'s in the simulated time series and plotted the area-normalized histograms for times between consecutive events (**Figure 1A**) and times between every second event (**Figure 1B**). We normalized histogram bin counts and square roots of bin counts by the total area under the histogram. Note that the square roots of the bin counts represent errors in bin counts. The probability density function obtained after area-normalization had a total area of 1. We fitted an exponential distribution, $a \times \exp(b \times x)$, to the distribution of times between consecutive events. The best-fit parameter for the decay rate $k = 51.0 \pm 0.1\,\text{s}^{-1}$ (mean ± standard error of the mean, SE) reasonably agrees with the average $52.6 \pm 0.1\,\text{s}^{-1}$ (mean ± SE) of $\Delta t_1$'s and the input parameter ($50\,\text{s}^{-1}$) for the simulation. We fitted a gamma distribution, $((b^a)/\Gamma(a)) \times x^{(a-1)} \times \exp(-b \times x)$, to the distribution of times between every second event (**Figure 1B**). The best-fitted shape and rate parameters are $2.11 \pm 0.01$ (mean ± SE) and $0.05 \pm 0.01$ (mean ± SE), respectively. With $dt = 1\,\text{ms}$, $10^7$ time points translate into $10^4\,\text{s}$. We divided the time series into $10^4$ blocks of time, each of 1 s duration, and counted the number of events (1's). As a result, we obtained a list of $10^4$ numbers of events per second. We plotted the area-normalized histogram and fitted a Gaussian distribution $a \times \exp(-((x-b)/c)^2)$ (**Figure 1C**). The best-fit parameter for the center of the Gaussian fit is $50.3 \pm 0.2\,\text{s}^{-1}$, which is close to the simulated rate $50\,\text{s}^{-1}$. For fits, we quantified the goodness-of-fit using a reduced chi-squared test. We calculated the reduced chi-squared value using the following equation (23),

$$\chi^2 = \frac{1}{N-M} \sum_{i=1}^{N} \frac{\left(x^i_{data} - x^i_{sim/fit}\right)^2}{\sigma_i^2}$$

where $N$ is the total number of bins, $M$ is the number of parameters, and $\sigma_i$ is the error at each histogram bin. For a perfect match, $\chi^2$ is zero, and therefore, a lower value indicates a good fit. We compared the $\chi^2$ values for different fit functions and chose the fit function with the lowest $\chi^2$ value. The $\chi^2$ values for the exponential (**Figure 1A**), the gamma (**Figure 1B**), and the Gaussian fits (**Figure 1C**) are $1.4304 \times 10^{-4}$ $1.0186 \times 10^{-4}$, and $0.0031$, respectively.

To investigate how the distributions change with the rate, we repeated the simulations and analyses for Poisson processes with rates $5\,\text{s}^{-1}$ (**Figure 1D-1F**) and $0.5\,\text{s}^{-1}$ (**Figure 1G-1I**). The distributions of times between consecutive events (**Figure 1D and 1G**) and times between every second event (**Figure 1E and 1H**) do not change shapes and recover the input rates reasonably well. However, the distributions of events per second change shape as the underlying rates of Poisson process change. The distributions are Gaussian (**Figure 1C**), Poisson (**Figure 1F**), and exponential (**Figure 1I**) for 50 s$^{-1}$, 5 s$^{-1}$, and 0.5 s$^{-1}$, respectively. For the Poisson distribution fit, we used the equation $(a^x / x!) \times exp(-a)$. The best-fit value for the Poisson distribution fit (**Figure 1F**) is $a = 4.99 \pm 0.02$, whereas the best-fit value of the decay rate for the exponential fit is $0.86 \pm 0.21$. In other words, a Poisson process can lead to different distributions depending on the rate for the events in a user-defined block of time. The agreement between the expected and observed consequences suggests that we simulated the Poisson process correctly.

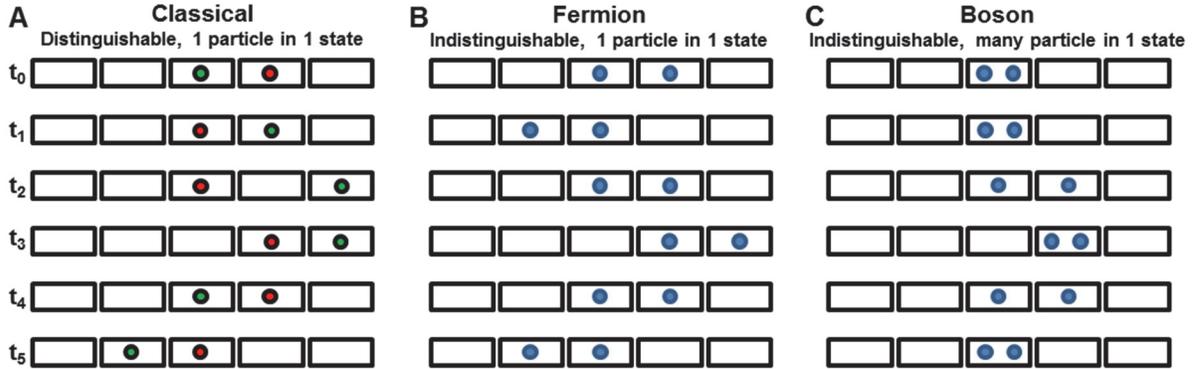

**Figure 2. Modeling arrow of time as an emergent property of random walk.** (**A**) Example of the time evolution of 2 classical particles in 5 states. The two particles return to the same initial state at $t_4$. Therefore, the time between consecutive recurrences $\Delta T1 = t_4 - t_0$. (**B**) Example of the time evolution of 2 Fermions in 5 boxes. The two particles return to the same initial state at $t_2$ and $t_4$. Therefore, there are two consecutive recurrences in this example, $\Delta T1 = t_2 - t_0$ and $t_4 - t_0$. (**C**) Example of the time evolution of 2 Bosons in 5 states. The two particles return to the same initial state at $t_1$, $t_3$, and $t_5$. Therefore, there are three consecutive recurrences in this example, $\Delta T1 = t_1 - t_0$, $t_3 - t_0$, and $t_5 - t_0$.

**A Poisson process can model the arrow of time.** We considered a system of particles undergoing a Poisson process to show that the arrow of time emerges naturally as the system size grows. To this end, we considered randomly walking and diffusing particles that can move to left or right with a constant probability of 0.5 at each time step. A one-dimensional (1D) random walk is also a Poisson process because it has a constant probability of going left or right at each time step.

We considered a 1D grid of $S$ boxes, where each box represents a single-particle state. At $t_0$, we randomly distributed $N$ particles in single-particle states represented by the boxes. During the initial assignment and subsequent random walk, we considered three types of particles. Classical particles are distinguishable, and only one particle can occupy a box. For Fermions, particles are indistinguishable, and only one particle can fill a box. Bosons are also indistinguishable, but many Bosons can occupy a box. Whenever the box occupancy violated these rules, we did not move to the next time step. We continued generating random numbers until particles filled the boxes according to the laws of occupancy. We considered the particles to be non-interacting.

After the initial random assignment in boxes, each particle underwent a 1D random walk to the left or right. At each time step, a particle had 1/3 probability of moving to the left or stay in the same box or move to the right. If a particle reached the boundary, it reflected and mimicked random walk in confined space. Particles could only move to the neighboring boxes at each time step. If two or more particles were to occupy the same state for classical and Fermion at a time step, we continued random number generation until the particles filled different boxes at that time step before moving to the next time step.

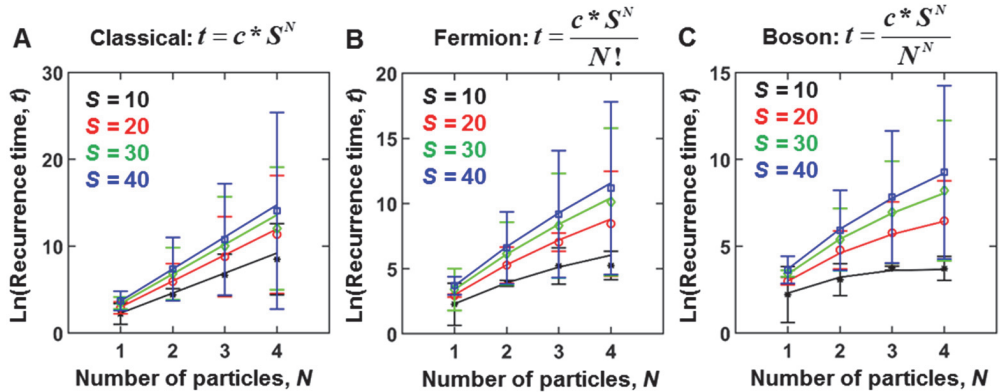

**Figure 3. Recurrence time depends on particle statistics.** A grid of size, $S$, represents the states that particles can occupy. Initially, a random number generator determines the locations of particles. As the particles undergo a random walk, there is a probability that particles would come back to initial positions (recurrence). The symbols represent the mean and standard error of the mean calculated from 10000 recurrence times for each condition. (**A**), (**B**), and (**C**) recurrent time as a function of the number of particles obeying classical, Fermion, and Boson statistics, respectively. Symbols represent simulated data points, whereas solid lines represent theoretical recurrence times calculated using the expression for recurrence time, $t$, in each case. Recurrence time depends on particle statistics as well as relative sizes of $S$ and $N$. The constant c is the same for the three statistics.

**The arrow of time is an emergent property for a larger system. Figure 3** shows that the average number of steps for recurrence grows as the system size becomes large. The recurrence time depends on the particle statistics, the size of the grid, and the number of particles. Even for 4 particles, a system takes a considerably large number of steps before returning to the initial state. The expressions for recurrence times for the three statistics are in **Figure 3**. Our simple 1D model leads to the irreversible arrow of time for large systems even though the dynamics of individual particles are reversible. In higher dimensions, the recurrence time follows $t = c*S^{d*N}$ for classical particles, where $d$ is the dimension. Although the arrow of time emerges for the three statistics, recurrence time depends on the particle statistics. Classical particles have the longest recurrence times, whereas Bosons have the shortest recurrence times (**Figure 3**).

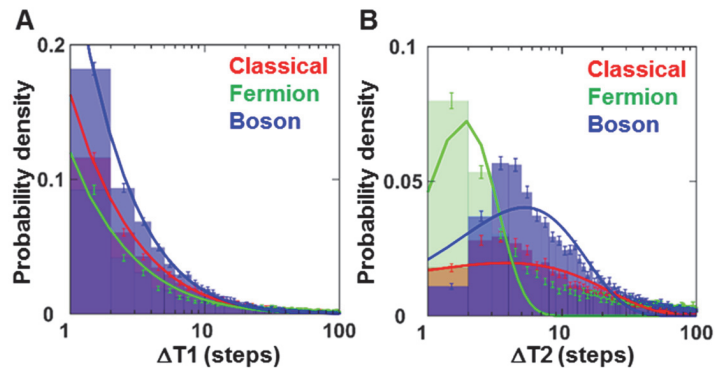

**Figure 4. Distribution of recurrence times.** (**A**) Area-normalized histograms of 10000 times ($\Delta T1$) between consecutive recurrences for $S=10$ and $N=2$. (**B**) Area-normalized histograms of 10000 times ($\Delta T2$) between every second recurrence. The solid lines represent the best fits. The error bars represent the square roots of the bin counts.

**A Poisson process does not describe the recurrence perfectly.** We argued that the underlying random walk of particles is a Poisson process. We investigated whether or not the recurrence itself is also a Poisson process by fitting exponential distributions to the histograms of consecutive recurrence times. A stretched exponential (24), $a \times exp(-(b \times x)^c)$, fits the distributions (**Figure 4A**). The $\chi^2$ values are 0.0012 (classical), 0.0018 (Fermion), and $6.2058 \times 10^{-4}$ (Boson). The best-fit values of the exponent $c$ are $0.17 \pm 0.01$ (classical), $0.15 \pm 0.02$ (Fermion), and $0.18 \pm 0.01$ (Boson). We fitted the distributions of times between every second event to a gamma distribution (**Figure 4B**). A gamma distribution does not fit the distributions of times between every second recurrence well. In other words, a Poisson process cannot describe the recurrence even though the underlying process of 1D diffusion is a Poisson process.

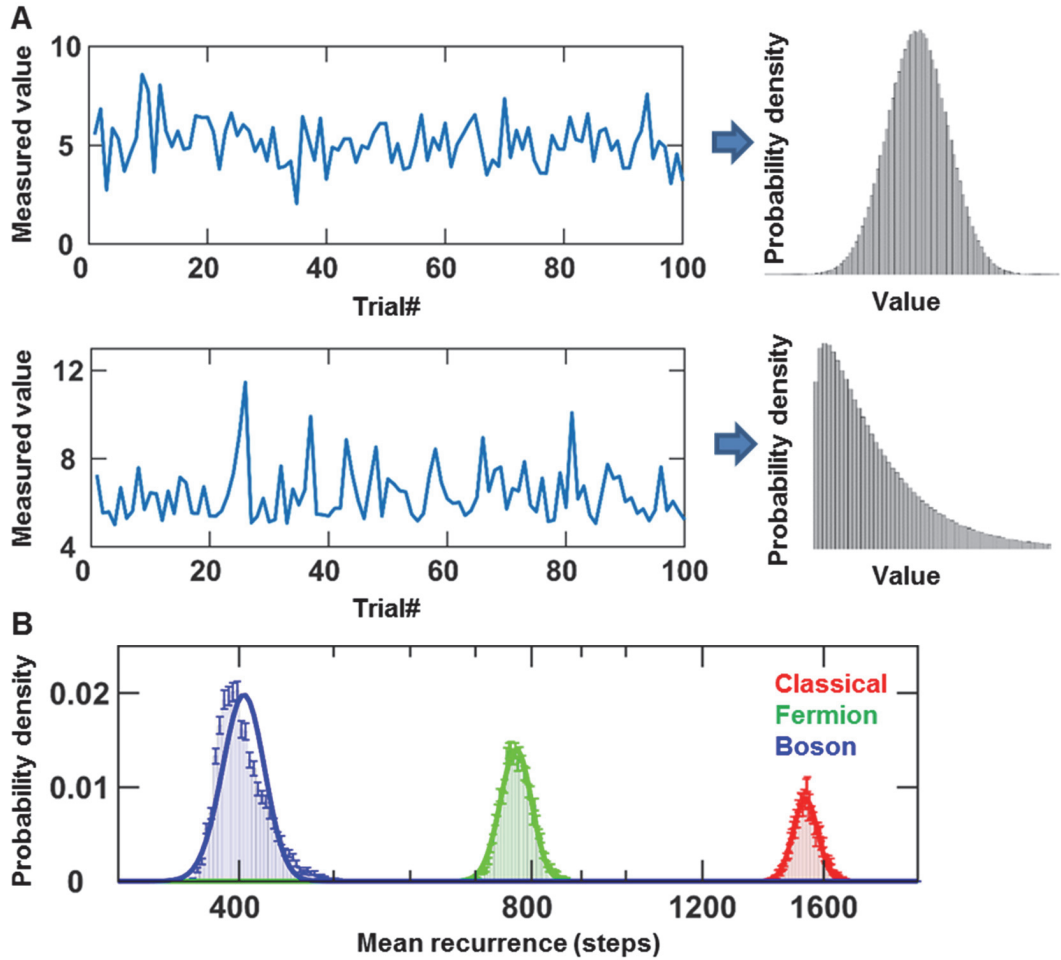

**Figure 5. Underlying distributions of the mean recurrence time.** (**A**) Top row: Simulated fluctuations around a value of 5 with an underlying Gaussian distribution. The mean of 1 million simulated noisy values is 5, which agrees with the input value. Bottom row: Simulated fluctuations with an underlying gamma distribution with both the shape and rate parameters set at 1.2. The mean of 1 million simulated noisy values is 6.44, which does not agree with the real value. (**B**) The distribution of the means of repeats for $S=40$ and $N=2$. Each repeat calculates the average of 10000 $\Delta T1$s. Classical particles (4481 repeats) and Fermions (5000 repeats) fit to a Gaussian-distributed mean, whereas Bosons (5000 repeats) fit better to a Poisson distribution. The solid lines represent the best fits. The error bars represent the square roots of the bin counts.

**The recurrence for Bosons leads to non-Gaussian distribution.** In science, we repeat experiments and simulations to define the accuracy and precision. For accuracy, we need standards to compare with the results. For recurrence times, the expressions for the recurrence times (**Figure 3**) provide the values that we can use as standards and compare them with the simulated results. For precision, we need to repeat experiments or simulations and quantify how well we can determine the mean. For both accuracy and precision, we often assume that the underlying fluctuations of values are Gaussian-distributed around the accurate value and calculate the mean and standard deviation (or standard error of the mean). When the underlying variations are Gaussian-distributed, an averaging of repeats provides the accurate value of the observable, and the width of the distribution provides the precision (**Figure 5A**, top panel). However, when the underlying fluctuations are gamma-distributed, an averaging does not give the accurate value (**Figure 5A**, bottom panel). To determine the underlying distributions for the recurrence times, we repeated simulations. For each repeat of simulation, we calculated the mean of 10000 recurrences for $S$=40 and $N$=2. **Figure 5B** shows that the underlying distributions for classical particles and Fermions are well-described by Gaussians. Still, the underlying distribution for Bosons is non-Gaussian and fits better with a Poisson distribution. In other words, the mean recurrence time and fluctuations around the mean depend on the particle statistics.

In summary, we have shown that the arrow of time naturally emerges as a consequence of a large system for classical particles, Fermions, and Bosons. As the number of particles in a system becomes large, the recurrence time grows, leading to the arrow of time and explains how dynamics with time-reversal symmetry can lead to the irreversibility of the arrow of time. Interestingly, the recurrence time depends on particle statistics. A system of Bosons returns to its initial configuration faster than Fermions and classical particles. A Poisson process describes the recurrence for Bosons better than Fermions and classical particles. It is interesting to think about the shorter recurrence time for Bosons in connection with the fact that Bosons came into being before Fermions and classical particles. We have shown that fluctuations around the mean recurrence time follow a Poisson distribution for Bosons and a Gaussian distribution for Fermions and classical particles. The concept of arrow of time may benefit students to learn some concepts of statistical mechanics as well as data analysis and programming.

**Acknowledgments**
The authors acknowledge Chase Harms for participating in the initial stage of simulations and analyses. S.K.S. acknowledges the students in thermodynamics and statistical mechanics classes for enduring the homework related to the concepts of this paper.

**Author contributions**
S.K.S. conceived and designed the overall project. D.W., R.K.-S., and S.K.S. performed the simulations and analyses. S.K.S. wrote the manuscript. All authors read and edited the manuscript.

**Competing financial interests**
The authors declare no competing financial interests.